\documentclass[prb, preprint, amsmath,amssymb]{revtex4-1}
\usepackage{multirow}
\usepackage{graphicx}
\usepackage{bm}
\usepackage{color}
\usepackage[colorlinks=true,linkcolor=blue, citecolor= red, pdfborder={0 0 0}]{hyperref}

\def\etal{{\it et~al.}}

\def\s1{\sigma_1(\omega)}

\begin{document}

\title{\boldmath Anisotropy of the infrared-active phonon modes in the mono-domain state of tetragonal SrTiO$_3$ (110)\unboldmath}

\author{M. Yazdi-Rizi}%
\email{meghdad.yazdi@unifr.ch}
 \affiliation{University of Fribourg, Department of Physics and Fribourg Center for Nanomaterials,
Chemin du Mus\'{e}e 3, CH-1700 Fribourg, Switzerland}
\author{P. Marsik}
\affiliation{University of Fribourg, Department of Physics and Fribourg Center for Nanomaterials,
Chemin du Mus\'{e}e 3, CH-1700 Fribourg, Switzerland}
\author{B. P. P. Mallett}
\affiliation{University of Fribourg, Department of Physics and Fribourg Center for Nanomaterials,
Chemin du Mus\'{e}e 3, CH-1700 Fribourg, Switzerland}
\author{C. Bernhard}%
\email{christian.bernhard@unifr.ch}
 \affiliation{University of Fribourg, Department of Physics and Fribourg Center for Nanomaterials,
Chemin du Mus\'{e}e 3, CH-1700 Fribourg, Switzerland}

\date{\today}

\begin{abstract}

With infrared and terahertz ellipsometry we investigated the anisotropy of the infrared active phonon modes in SrTiO$_3$ (110) single crystals in the tetragonal state below the so-called antiferrodistortive transition at T$^*$ = 105 K. In particular, we show that the anisotropy of the oscillator strength of the so-called R-mode, which becomes weakly infrared active below T$^*$, is a valuable indicator for the orientation of the structural domains. Our results reveal that a mono-domain state with the tetragonal axis (c-axis) parallel to the [001] direction can be created by applying a moderate uniaxial stress of about 2.3 MPa along the [1-10] direction (with a mechanical clamp). The resulting, intrinsic splitting of the infrared-active phonon modes is reported.

\end{abstract}



\maketitle


\section{Introduction}
SrTiO$_3$ (STO) is widely used as a substrate for growing thin films and multilayers of various complex oxides. Furthermore, the interface of STO with other insulating oxide perovskites, like LaAlO$_3$ (LAO), has become a subject of great interest since it was found that it can host a two-dimensional electron gas with a very high mobility \cite{Ohtomo2004, Thiel2006} and even superconductivity \cite{Reyren2007} and ferromagnetism \cite{Bell2009} at low temperature. These developments have renewed the interest in the structural and electronic properties of bulk STO.

With decreasing temperature, STO undergoes a series of structural phase transitions which break several symmetries \cite{Sulpizio2014}. The first, so-called anitferrodistortive transition at T$^*$ = 105 K from a cubic to a tetragonal symmetry involves an antiphase rotation of the neighboring TiO$_6$ octahedra around the c-axis, as shown in Fig.\ref{fig1}(a). It gives rise to a doubling of the unit cell along all three directions and a slightly larger lattice parameter in the direction of the rotation axis (c-axis) as compared to the perpendicular ones (a-axis) with a ratio of $c/a\approx$ 1.0015 at low temperature \cite{Loetzsch2010}. There are further structural transitions into an orthorhombic state around 65 K and a rhombohedral one at 37 K \cite{Lytle1964}. Below about 25 K STO enters a so-called incipient ferroelectric or quantum paraelectric state \cite{Zhong1996} for which a ferroelectric order is only prohibited by the ionic quantum fluctuations \cite{Muller1979}. Finally, at T $<$ 10 K it has been reported that STO can exhibit a piezoelectric response \cite{Grupp1997}.

The structural and electronic properties of STO are rather sensitive to strain and defects. In the tetragonal state below T$^*$ = 105 K this typically results in a structural poly-domain state with different orientations of the tetragonal axis (c-axis) \cite{Fleury1968, Sawaguchi1963}. These poly-domain structures and their microscopic imprint on the physical properties of STO and heterostructures in which STO serves as a substrate are well documented \cite{Hoppler2008}, \cite{Kalisky2013, Honig2013}. A preferred orientation of these structural domains can be achieved by applying uniaxial pressure \cite{Chang1972, Muller1970} or large electric fields \cite{vanMechelen2010thesis}. Furthermore, it was shown that the direction of the surface cut of the STO crystals can strongly affect the orientation of these structural domains. Whereas for STO with a (001) surface one typically finds a poly-domain state for which the possible orientations of the tetragonal axis appear with nearly equal probablity, for STO with a (110) surface the tetragonal axis tends to be preferentially aligned along the [001] direction\cite{Muller1970, Chang1970}. The latter may well be related to the anisotropy of the (110) surface \cite{Wang2014, Annadi2013}. Furthermore, the surface of STO (110) is polar and thus tends to undergo a reconstruction \cite{Bottin2005, Eglitis2008, Enterkin2010}. 

An important role of the surface in the antiferrodistortive transition is also suggested by the finding that in the vicinity of the sample surface this transition can occur at considerably higher temperatures (with T$\leq$ 150 K) than in the bulk (with T$^*$ = 105 K). The details have been shown to strongly depend on the residual strain that arises from the surface preparation, e.g. by lapping, grinding or cutting \cite{Chrosch1998} or the strain due to a thin film on top of the STO substrate \cite{Loetzsch2010}. This raises the question whether in thick STO (110) single crystals the domain alignment is limited to the near surface region \cite{Chrosch1998} or also occurs in the bulk. 
In the following we use the anisotropy of the infrared-active phonon modes, in particular, of the so-called R-mode which becomes infrared active below T$^*$ = 105 K, to study the structural domain formation in STO (110) crystals with a thickness of up to 1 mm. We find that the pristine STO (110) crystals are already partially detwinned with the tetragonal axis oriented along the [001] direction. Furthermore, we show that a relatively small uniaxial stress of about 2.3 MPa is sufficient to obtain a true mono-domain state. Notably, this mono-domain state is found to persist after the samples has been warmed to room temperature and cooled again without the uni-directional stress. Finally, we have also determined the anisotropy of the other three infrared-active phonon modes which is rather weak and in agreement with the small anisotropy ratio of $c/a\approx$ 1.0015 \cite{Loetzsch2010}.

\section{Experimental details}

The infrared ellipsometry measurements have been performed with a home-built setup that is equipped with a He flow cryostat and attached to a Bruker 113V Fast Fourier spectrometer as described in Ref. \cite{Bernhard2004}. The data have been taken at different temperatures in rotating analyzer mode, with and without a static compensator based on a ZnSe prism. The THz optical response has been measured with a home-built time-domain THz ellipsometer that is described in Ref. \cite{Marsik2016}. For both ellipsometers the angle of incidence of the light was set to 75 degrees.

For anisotropic samples the ellipsometric measurements are predominantly sensitive to the response along the plane of incidence of the reflected light \cite{Aspnes1980}. Accordingly, for the present STO (110) crystals, which in the tetragonal state exhibit an uniaxial anisotropy with the [001] symmetry axis parallel to the surface, we measured the dielectric response functions along the [001] and [1-10] directions by rotating the sample by appropriate angles around the surface normal. The optical conductivity data for the two orientations as shown in this manuscript have been obtained by directly calculating the pseudodielectric function $<\varepsilon>$ from the measured ellipsometric angles $\Psi$ and $\Delta$, where $r_p/r_s=tan(\Psi)e^{i\Delta}$ and $r_p$ and $r_s$ are the Fresnel reflection coefficients for p- and s- polarised light. During all the optical measurements special care was taken to avoid photo-doping effects by shielding the sample against visible and UV light \cite{Kozuka2007}. To apply a uniaxial stress of about 2.3 MPa, either along the [001] or the [1-10] direction, we mounted the sample in a spring-loaded clamp and cooled it slowly to low temperature. For the stress-free measurements the sample was glued with silver paint to a different sample holder. The STO (110) single crystals were purchased from SurfaceNet and had dimensions of 10$\times$10$\times$1 mm$^3$, 10$\times$10$\times$0.5 mm$^3$, 5$\times$5$\times$1 mm$^3$ and 5$\times$5$\times$1 mm$^3$, respectively.

\section{Experimental Results}

In the cubic state at T $>$ T$^*$ STO has three infrared-active transversal optical (TO) phonon modes that are triply degenerate \cite{Hlinka2006}, \cite{Trautmann2004}. At the lowest frequency is the so-called soft-mode (or Slater mode) with an eigenfrequency of $\omega_0\approx$ 95 cm$^{-1}$ at 300 K which decreases to about 15 cm$^{-1}$ at 10 K. It involves the displacement of the Ti ions against the surrounding oxygen octahedron and is at the heart of the quantum paraelectric behavior of STO. The phonon at 170 cm$^{-1}$ is the so-called external mode (or Last mode) due to the displacement of the Sr ions against the TiO$_6$ octahedra. At the highest energy of about 545 cm$^{-1}$ is the so-called stretching mode (or Axe-mode) which stretches the oxygen octahedra.

\begin{figure}[!h]
\begin{center}
\includegraphics[width=1\textwidth]{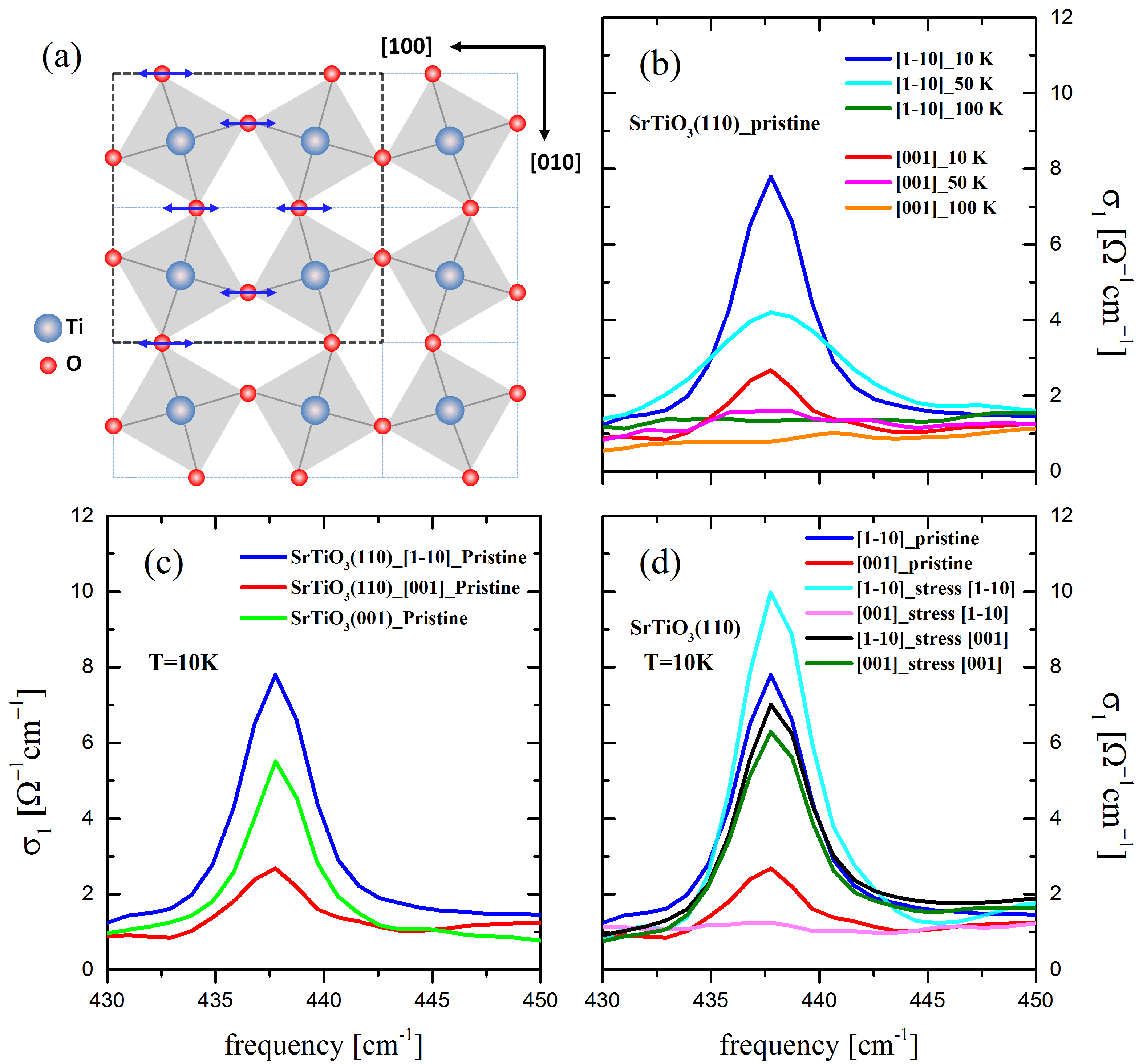}
\caption{(a) Sketch of the tetragonal structure of STO below T$^*$ = 105 K with the antiphase rotation of the TiO$_6$ octahedra around the tetragonal axis (c-axis). (b) Spectra of the real part of the optical conductivity of pristine STO (110) showing the so-called R-mode along the [1-10] and [001] directions at dixfferent temperatures below T$^*$ = 105 K. (c) Comparison of the R-mode at T=10 K in pristine STO (001) and STO (110) crystals that are lightly and heavily twinned, respectively. (d) The effect of a uniaxial stress of 2.3 MPa applied along the [1-10] and [001] directions on the spectra of the R-mode along [1 10] and [001], respectively.}
\label{fig1}
\end{center}
\end{figure}

In the tetragonal state below T$^*$ = 105 K an additional mode becomes weakly infrared-active due to the antiphase rotation of the neighboring oxygen octahedra, as sketched in Fig. \ref{fig1}(a). It is the so-called R mode at 438 cm$^{-1}$ \cite{Lytle1964} that arises from the doubling of the unit cell and the subsequent back-folding of the phonon modes from the boundary (the R-point) to the center of the Brilluoin zone \cite{Fleury1968, Petzelt2001}. The oscillator strength of this R-mode is proportional to the magnitude of the antiphase rotation of the TiO$_6$ octahedra. Furthermore, its strenght exhibits a characterisitc dependence on the polarization direction of the infrared light with respect to the tetragonal axis. It is maximal if the polarization is perpendicular to the rotation axis of the oxygen octahedra (to the tetragonal axis) and it vanishes when it is parallel. The large sensitivity of the R-mode to such structural changes has already been demonstrated in Ref. \cite{Rossle2013prl}.

In the following we show that this anisotropy of the oscillator strength of the R-mode can be used to determine the orientation of the tetragonal axis in STO single crystals and its variation in a structural multi-domain state. Figure \ref{fig1}(b) shows the temperature and polarization dependence of the R-mode of a pristine STO (110) single crystal in terms of the real part of the optical conductivity. It confirms that the oscillator strength of the R-mode is strongly anisotropic, i.e. it is almost four times larger along the [1-10] direction than along [001]. Figure \ref{fig1}(c) shows a comparison of the R-modes in pristine STO (110) and STO (001) single crystals. For the latter the oscillator strenght is isotropic (this has been confirmed for several STO (001) crystals) and its value is intermediate between the ones along [001] and [1-10] in STO (110). This agrees with previous reports that in STO (001) the domains are almost randomly oriented (the crystal is fully twinned) whereas in STO (110) the domains are preferentially oriented (the crystal is partially detwinned) with the tetragonal axis along the [001] direction. A corresponding behavior has been observed for all the STO (110) crystals in their pristine state.

Next, in Fig. \ref{fig1}(d) we show that a fully detwinned mono-domain state can be induced in STO (110) by applying a moderate uniaxial stress along the [1-10] direction. This stress gives rise to a small compression of the Ti-O bonds along [1-10] and, likewise, a small expansion along [001] and [110] that provides an additional incentive for the longer c-axis to be oriented along the [001] direction. For these measurements the STO (110) sample of dimensions 10$\times$10$\times$0.5 mm$^3$ was mounted in a spring-loaded holder that applies a uniaxial stress of about 2.3 MPa during the cooling from room temperature to 10 K (at a rate of 3 K/min) and the subsequent optical measurement. Figure \ref{fig1}(d) confirms that the R-mode is now essentially absent along the [001] direction, i.e. its oscillator strength is zero within the accuracy of the measurement. The oscillator strength along the [1-10] direction exhibits a corresponding increase as compared to the pristine state. The absence of the R-mode for the polarization along [001] suggests that the STO (110) crystal is in a mono-domain state with the tetragonal axis parallel to the [001] direction. 
We also found that the STO (110) crystal has a memory of this mono-domain state that was almost perfectly maintained after the sample was warmed to room temperature and slowly cooled again to 10 K without the uniaxial stress. Such a memory of the mono-domain state was observed for all the STO (110) samples. 
To erase this memory of the mono domain state, it was necessary to heat the sample to higher temperatures of T$\geq$410 K. Likewise, the twinning of the STO (110) crystal can be restored, and even strongly enhanced as compared to the pristine state, if the stress is applied along the [001] direction. Figure \ref{fig1}(d) confirms that this yields nearly equal oscillator strengths for the R-mode along the [001] and [1-10] directions. Note that for the STO (001) crystals we found that such a uniaxial pressure does not have a corresponding effect on the directional dependence of the oscillator strength of the R-mode and thus on the degree of the twinning below T$^*$.

\begin{figure}[!h]
\begin{center}
\includegraphics[width=1\textwidth]{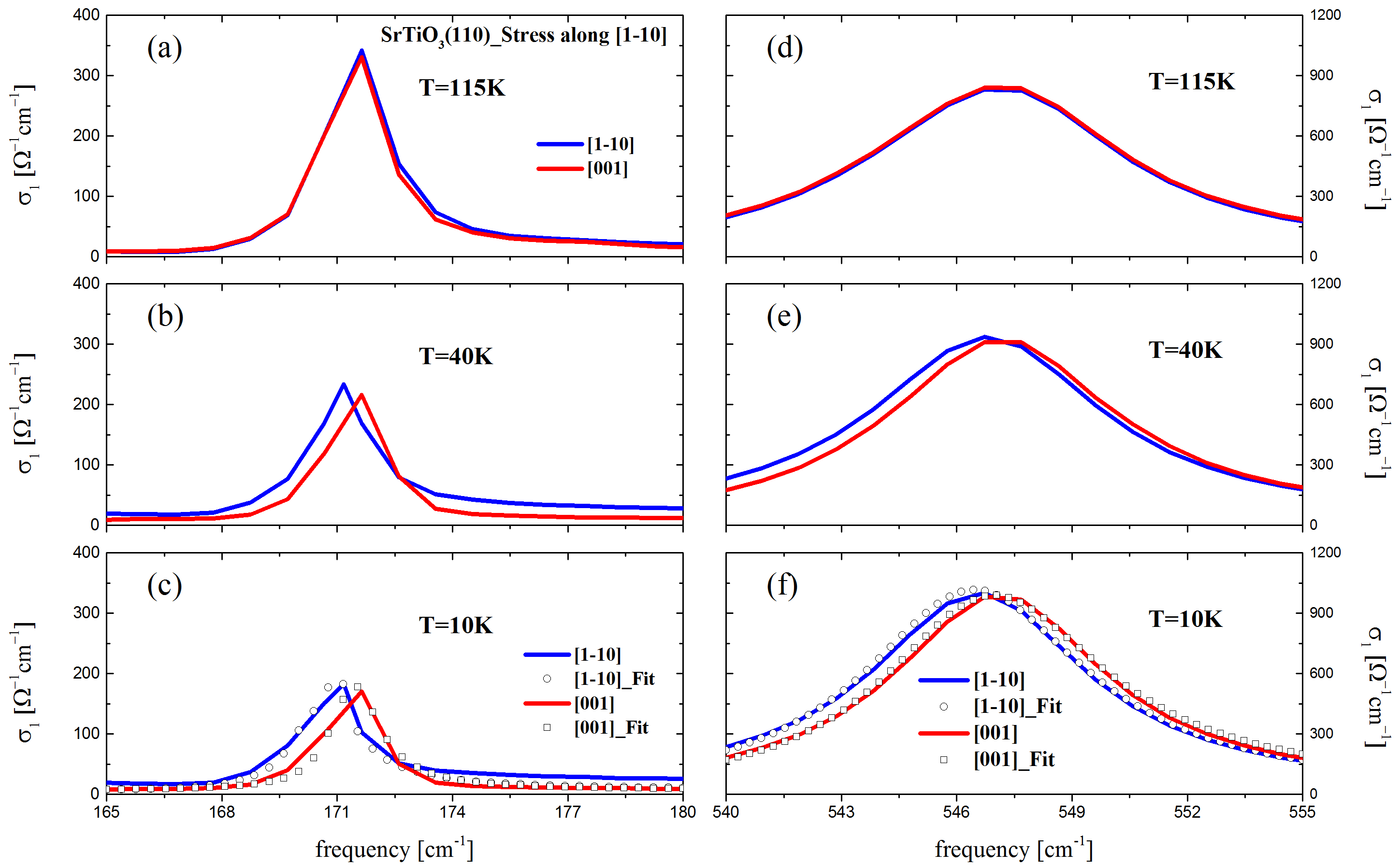}
\caption{Anisotropy of the phonon modes at 170 and 545 cm$^{-1}$ along the [1-10] and [001] directions for STO (100). The crystal was cooled under a uniaxial stress of 2.3 MPa that was applied along the [1-10] direction such that it is in a mono-domain state with the tetragonal axis along [001]. (a) and (d) Spectra of the optical conductivity in the cubic state at T=115 K $>$ T$^*$ for which the phonon modes are isotropic. (c) and (f) Corresponding spectra at 10 K where the phonon modes exhibit a small, yet clearly resolved splitting. This splitting is well reproduced by the fits with a so-called coupled phonon model \cite{Humlicek2000} that are shown by the open circles (the parameters are listed in Table \ref{table1}).}
\label{fig2}
\end{center}
\end{figure}

The possibility to obtain STO (110) crystals with a mono-domain tetragonal state below T$^*$ = 105 K was used to study the intrinsic anisotropy of the other IR-active phonon modes. Figure \ref{fig2} shows the data of the STO (110) crystal in the stressed state for the external mode at 170 cm$^{-1}$ and the stretching  mode at 545 cm$^{-1}$ in terms of the optical conductivity measured along the [001] and [1-10] directions. In the cubic state at T=115 K $>$ T$^*$ = 105 K, there is no sign of an anisotropy of these phonon modes, i.e. the conductivity in the [001] and [1-10] directions agrees within the noise level (see Figs. \ref{fig2}(a) and \ref{fig2}(d)). A small, yet clearly resolved anisotropy appears only in the tetragonal state at T $<$ T$^*$ = 105 K where both TO modes become slightly harder along the [001] direction and softer along [1-10]. This agrees with the expectation that the tilting of the Ti-O bonds in the plane perpendicular to the rotation axis (tetragonal axis) leads to a softening of the phonon modes (see e.g. \cite{Yamanaka2000}). 
To quantify the anisotropy, we parameterized the phonon modes using a modified Lorentzian model, the so-called coupled-phonon model \cite{Humlicek2000} which yields the following expression of the complex dielectric function:
\begin{equation}
\varepsilon(\omega)=\varepsilon_{1}(\omega)+i\varepsilon_{2}(\omega)=\varepsilon_{\infty}(\omega)+\sum\limits_{j=1}^{K}S_{j}\frac{\omega^{2}_{j}-i\omega\sigma_{j}}{\omega^{2}_{j}-\omega^{2}-i\omega\gamma_{j}}
\end{equation}
This model accounts for phonon modes that are asymmetric due to disorder or anharmonicity effects, as is required for a good descritpion of the phonon modes in STO \cite{Humlicek2000, Fedorov1998}. It contains the usual paramters, like the broadening, $\gamma_{j}$, the resonance frequency, $\omega_{j}$, and the oscillator strength, $S_{j}$, of the jth phonon mode. The coupling between these modes arises from the extra term $i\omega\sigma_{j}$ in the numerator that is subject to the condition $\sum_{j=1}^{K}S_{j}\sigma_{j}=0$ which ensures that the Kramers-Kronig consistency is maintained\cite{Humlicek2000}. The background term $\varepsilon_{\infty}$ accounts for the frequency independent contribution from the excitations at higher energies.
The best fits are shown in Figure \ref{fig2} by the solid symbols and describe the phonon modes rather well (except for a small and only weakly frequency dependent offset to the conductivity along [001] that most likely arises from a spurious reflection on the clamp). The obtained parameters for the eigenfrequency, the broadening and the oscillator strengths of the phonon modes are listed in Table \ref{table1}. The parameters for $\sigma_{j}$ and $\varepsilon_{\infty}$ have been adopted from Ref. \cite{Rossle2013thesis, Rossle2013prb} and the ones for the soft-mode have been obtained from the THz data as described further below.

The farily weak splitting of the eigenfrequencies of the external mode at 170 cm$^{-1}$ and the stretching mode at 545 cm$^{-1}$ of about 1 cm$^{-1}$ and 0.7 cm$^{-1}$, respectively, is in good agreement with the small anisotropy ratio of the lattice parameters of $c/a\approx$ 1.0015 \cite{Loetzsch2010}. Note that the contribution of the applied stress (2.3 MPa) to the splitting of the phonon modes is expected to be significantly smaller, i.e. less than 0.1 cm$^{-1}$. This estimate is based on the pressure dependence of the Raman modes \cite{Guennuo2010} as well as the pressure- and temperature dependence of the lattice parameter \cite{Guennuo2010, Loetzsch2010} and the temperature dependence of the eigenfrequency of the phonon modes of STO \cite{Rossle2013thesis}. 

\begin{figure}[!h]
\begin{center}
\includegraphics[width=.9\textwidth]{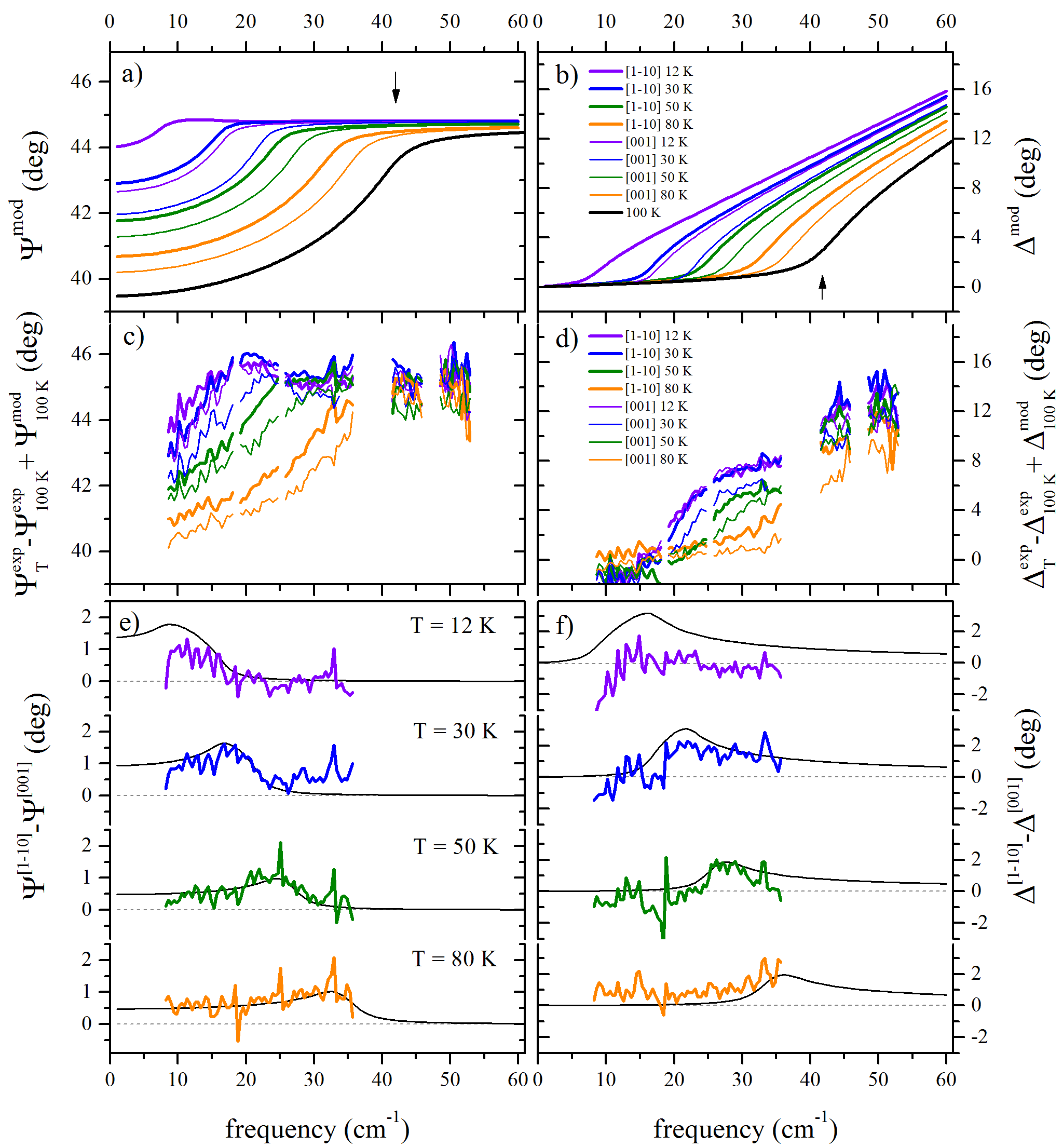}
\caption{Anisotropy of the soft phonon mode of STO (110) in the THz-range. (a, b) Calculated spectra of the ellipsometric angles $\Psi$ and $\Delta$ for the isotropic soft-mode response at 100 K and the anisotropic one at lower temperatures. The phonon parameters that we used are specified in the text. (c) and (d) Measured spectra of $\Psi$ and $\Delta$ for the orientations along [1-10] and [001] (with the sample clamped along [1-10]). To remove artifacts related to light scattered on the clamp, the plots show the difference spectra with respect to 100 K to which the calculated spectrum at 100 K (thick black lines in (a) and (b)) has been added. (e) and (f) Comparison of the calculated (black line) and the measured anisotropy of $\Psi$ and $\Delta$ between the [1-10] and [001] directions.}
\label{fig3}
\end{center}
\end{figure}

Figure \ref{fig3} shows the corresponding anisotropy of the soft mode in STO (110). Figures \ref{fig3}(a) and (b) show the calculated values of the ellipsometric angles $\Psi$ and $\Delta$ as obtained with an isotropic phonon model at 100 K (black line) and with an anisotropic model below 100 K (coloured lines). The phonon parameters used for these calculations have been obtained from Ref. \cite{Yamanaka2000, Fedorov1998, Rossle2013thesis}. In particular, we have adopted from Ref. \cite{Yamanaka2000} the anisotropy of the TO-frequency at 10 K which amounts to $\omega_{TO}$ = 8 and 17 cm$^{-1}$ along [1-10] and [001], respectively. This splitting of the soft mode decreases with increasing temperature and vanishes at T$^*$. The spectral weight of the soft-mode is assumed to be temperature independent and identical for both orientations. This assumption is consistent with the far-infrared response and leads to a virtually constant value of the corresponding LO frequency, as reported in Ref.\cite{Fedorov1998} and also shown in Fig. \ref{fig5}.

The broadening, $\gamma$, was set to 6, 5, 4, 3 and 3 cm$^{-1}$ at 100, 80, 50, 30 and 10 K, respectively, in agreement with Refs. \cite{Yamanaka2000, Fedorov1998, Rossle2013thesis}. Figure \ref{fig3}(a) and (b) show that the spectra of $\Psi$ and $\Delta$ exhibit pronounced kinks in the vicinity of $\omega_{TO}$ that are marked by solid arrows for the spectrum at 100 K for which $\omega_{TO}$ = 41.8 cm$^{-1}$.

Figures \ref{fig3}(c) and (d) show the experimental spectra of $\Psi$ and $\Delta$ as obtained with the time-domain THz ellipsometer. Due to the rather large focal spot of the THz beam, the raw data are affected by stray light that is reflected from the clamp in which the sample is mounted. To suppress this experimental artifact, the data are presented in terms of the difference spectra of $\Psi$ and $\Delta$ with respect to the curves at 100 K to which the calculated spectrum at 100 K (solid lines in Fig. \ref{fig3}(a) and (b)) has been added.  The spectra provide clear evidence for a sizable anisotropy of the soft-mode below 100 K. The comparison between the calculated (black lines) and the measured anisotropy of $\Psi$ and $\Delta$ between the [1-10] and [001] directions is shown in Figs. \ref{fig3}(e) and (f). The fair agreement between the calculated and the measured data (except for a frequency independent vertical offset of some spectra that are likely due to an incomplete suppression of the stray light signal) confirms that the soft-mode develops a  sizable anisotropy below T$^*$.

\begin{figure}[!h]
\begin{center}
\includegraphics[width=1\textwidth]{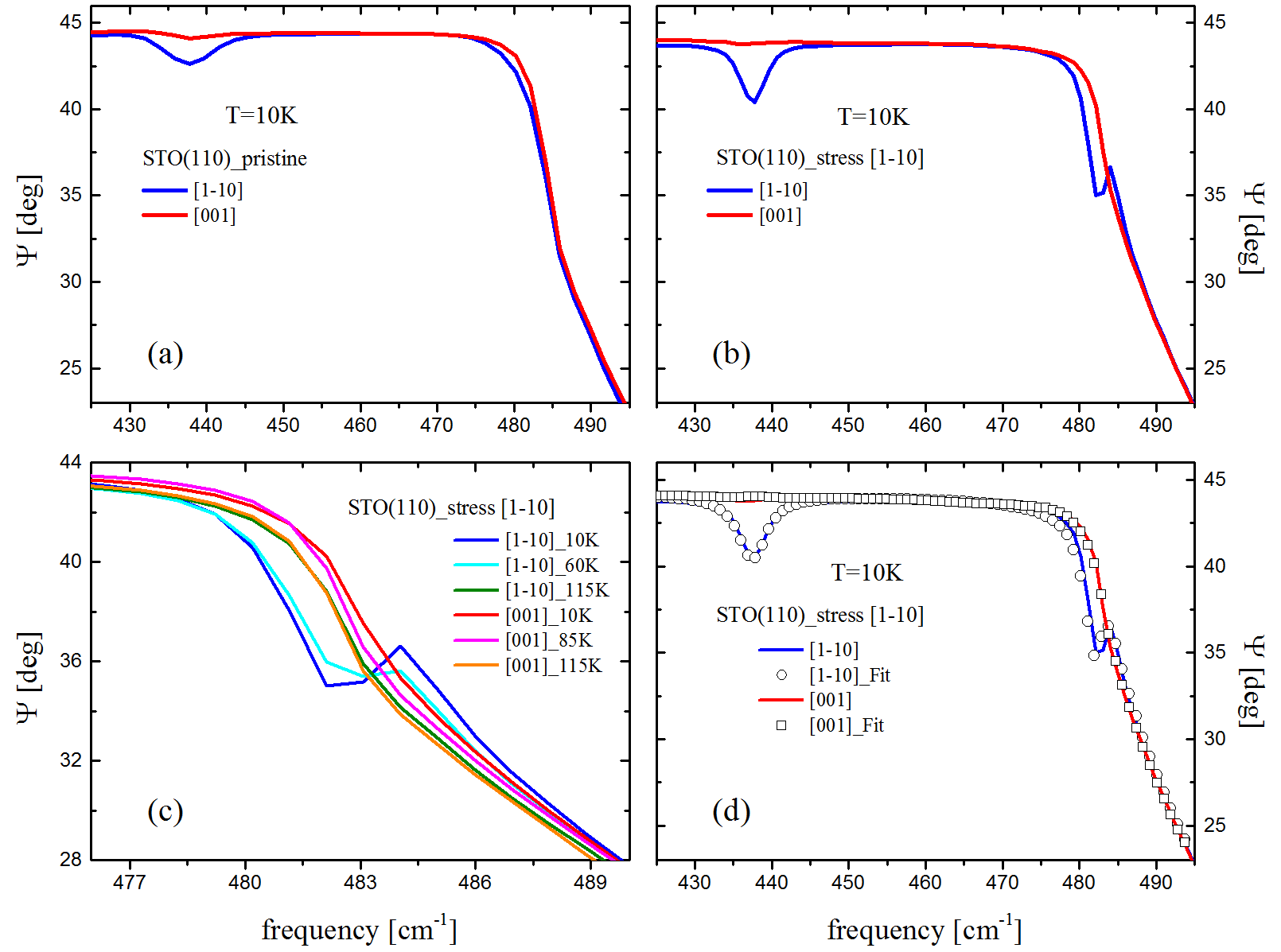}
\caption{Dip-feature in the spectrum of $\Psi$ in the vicinity of the LO-mode of STO (110) around 480 cm$^{-1}$ and its dependence on the structural domain state. Also shown is the R-mode around 438 cm$^{-1}$ from which the degree of detwinning can be deduced. (a) and (b) Data for STO (110) in pristine state and after cooling under uniaxial stress along [1-10] where the crystal is still partially twinned and in a mono-domain state, respectively. It shows that the dip-feature around 480 cm$^{-1}$ is more pronounced and sharper in the mono-domain state. (c) Evolution of the dip-feature in the mono-domain state as a function of temperature. (d) Comparison of the experimental data in the mono-domain state at 10 K (solid lines) with the fit using an anisotropic version of the coupled phonon model (open circles).}
\label{fig4}
\end{center}
\end{figure}

\begin{figure}[!h]
\begin{center}
\includegraphics[width=.8\textwidth]{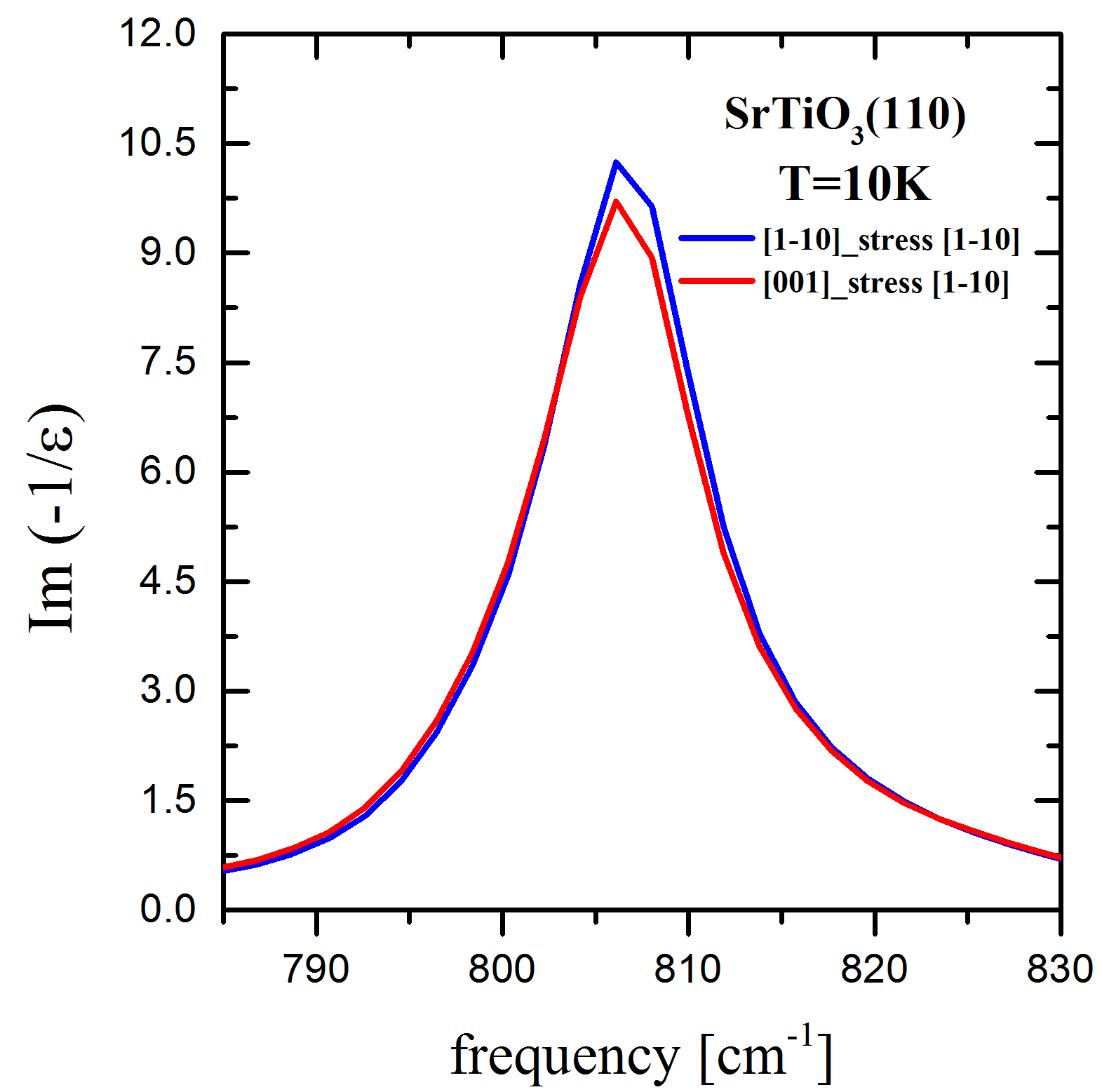}
\caption{Spectra of the loss function in the vicinity of the highest LO-phonon mode of STO (110) in the mono-domain state as obtained by cooling the sample under uniaxial stress of 2.3 MPa applied along the [1-10] direction. A very small anisotropy of the highest LO mode of less than 0.5 cm$^{-1}$ is evident between the responses along the [001] and [1-10] directions.}
\label{fig5}
\end{center}
\end{figure}

Next, we discuss an additional, sharp feature around 480 cm$^{-1}$ in the IR spectra of STO (110) that becomes very pronounced in the mono-domain state at low temperature. Figure \ref{fig4} shows the spectra of the ellipsometric angle $\Psi$ of STO (110) in the frequency range between 425 and 500 cm$^{-1}$ which includes the R-mode at 438 cm$^{-1}$ from which the degree of twinning can be deduced. The new feature develops around 480 cm$^{-1}$ very close to a longitudinal optical phonon mode frequency at which the real part of the dielectric function, $\varepsilon_{1}$, crosses zero. Figures \ref{fig4}(a) and \ref{fig4}(b) compare the spectra for the partially twinned pristine and the mono-domain state. In the mono-domain state, there is a rather sharp dip feature in the spectrum along [1-10] that is absent along [001]. In the pristine state this dip feature is much broader and therefore barely visible as a small difference between the spectra along [1-10] and [001]. Figure \ref{fig4}(c) shows that this dip feature at 480 cm$^{-1}$ in the mono-domain state along [1-10] vanishes when the temperature is increased above the structural transition at T$^*$ = 105 K where the spectra for [1-10] and [001] agree within the noise level.

In the following we show that this dip feature can be understood in terms of the anisotropy of the IR active phonon modes. Specifically, it results from a small splitting of the zero-crossings of $\varepsilon_{1}$ along the [1-10] and [001] directions that is mainly caused by the splitting of soft mode along the in-plane directions. Figure \ref{fig4}(d) shows that this feature can be reproduced with a modified version of the coupled-phonon model \cite{Humlicek2000} which takes into account the anisotropy of the phonon modes along the [001] and the [1-10] directions. For the simulation shown by the solid symbols in Fig. \ref{fig4}(d) we used for the phonon modes at 170, 438 and 545 cm$^{-1}$ the eigenfrequency, oscillator strength and broadening as listed in Table \ref{table1} and the corresponding phonon asymmetries and value of $\varepsilon_{\infty}$ as reported in Refs. \cite{Rossle2013thesis, Rossle2013prb}. For the soft mode we used a splitting of the TO mode that is somewhat smaller than the one that was reported in Ref. \cite{Yamanaka2000} and used in describing the THz data in Fig. \ref{fig3}. The position of the highest LO-mode has been obtained from the data shown in Fig. \ref{fig5} as is discussed further below.

An alternative interpretation of this dip feature around 480 cm$^{-1}$ in terms of a plasmonic effect due to a so-called Berreman mode \cite{Berreman1963} which arises from mobile charge carriers that are confined to the surface of the sample can be discarded. Such a Berreman mode was observed at the interface of LAO/STO heterostructures where it gives rise to a pronounced dip feature at the highest LO phonon mode of STO around 865 cm$^{-1}$ \cite{Dubroka2010, Yazdi-Rizi2016}. Nevertheless, the present STO (110) sample does not have such a hetero-interface and, furthermore, a corresponding dip feature does not occur at the highest LO-edge near 865 cm$^{-1}$ where, according to the Berreman-mode scenario, it should be strongest.

Finally, Figure \ref{fig5} shows the highest LO mode of STO (110) in terms of the so-called loss-function. It confirms that this LO mode is centered around 806 cm$^{-1}$ and exhibits a very small anisotropy between the [1-10] and [001] directions of less than 0.5 cm$^{-1}$. This is expected since this LO mode contains a large contribution from the soft mode and $\omega_{TO}$ is the mean square value of the contributions that arise from the Coulomb-repulsion and the restoring force of the Ti-O bonds of which only the latter is affected by the lattice anisotropy.

\section{Summary}

In summary, with infrared ellipsometry we studied the anisotropy of the IR-active modes in STO (110) which develops below the cubic-to-tetragonal phase transition at T$^*$ = 105 K. We have shown that the tetragonal axis is preferentially oriented along the [001] direction. A partial orientation of the structural domains is already seen for the pristine STO (110) crystals and a relatively small stress of 2.3 MPa along the [1-10] direction is sufficient to induce a mono-domain state. The memory of this mono-domain state can be almost preserved after warming the sample to room temperature and cooling it again under stress-free conditions. We have also determined the weak anistropy of the infrared-active phonon modes at T=10 K which is in agreement with the small lattice parameter ratio of $c/a\approx$ 1.0015. The only exception concerns the so-called R-mode at 438 cm$^{-1}$ which only becomes infrared-active due to the antiphase rotation of the TiO$_6$ octahedra below T$^*$ = 105 K. Its oscillator strength exhibits a huge anisotropy and is maximal (zero) in the direction perpendicular (parallel) to the rotation axis (the tetragonal axis).

\begin{acknowledgments}
The work at the University of Fribourg has been supported by the Schweizerische Nationalfonds (SNF) through the grant No. 200020-153660. CB acknowledges fruitful discussions with Dominik Munzar. 
\end{acknowledgments}

\break

\begin {table}[!t]
\caption {Phonon parameters obtained from the best fit using the so-called coupled-phonon model of Ref. \cite{Humlicek2000}.} \label{table1} 
\begin{center}
\begin{tabular}{|cc|c|c|c|}
\cline{3-5}
\multicolumn{2}{c|}{\textit{STO (110) - STRESS [1-10]}} & External mode &   R-mode   & Stretching mode\\
\cline{1-5}
\multirow{2}{*}{Energy ($cm^{-1}$)} & [1-10]& 170.43& 437.8& 546.01  \\
                         & [001]& 171.40& -& 546.72  \\
\cline{1-5}
\multirow{2}{*}{Oscillator strength ($cm^{-1}$)} & [1-10]&0.72& 0.01& 1.45  \\
                         & [001]& 0.67& - & 1.37 \\
\cline{1-5}
\multirow{2}{*}{Broadening ($cm^{-1}$)} & [1-10]& 1.51& 4.50& 6.99  \\
                         & [001]& 1.51& -& 6.99 \\
\hline

\end{tabular}
\end{center}
\end {table}


\begin{thebibliography}{99} \markboth{Bibliography}{Bibliography}


\bibitem{Ohtomo2004}
A.A. Ohtomo \etal, Nature {\bf 427}, 423 (2004)

\bibitem{Thiel2006}
S. Thiel  \etal, Science {\bf 313}, 1942 (2006)

\bibitem{Reyren2007}
N. Reyren  \etal, Science {\bf 317}, 1196 (2007)

\bibitem{Bell2009}
C. Bell \etal, Phys. Rev. Lett. {\bf 103}, 226802 (2009)

\bibitem{Sulpizio2014}
J. A. Sulpizio \etal, Annu. Rev. Mater. Res. {\bf 44}, 117 (2014)

\bibitem{Loetzsch2010}
R. Loetzsch \etal, Appl. Phys. Lett. {\bf 96}, 071901 (2010)

\bibitem{Lytle1964}
F. W. Lytle, J. Appl. Phys. {\bf 35}, 2212 (1964)

\bibitem{Zhong1996}
W. Zhong \etal, Phys. Rev. B {\bf 53}, 5047 (1996)

\bibitem{Muller1979}
K. A. M\"uller \etal, Phys. Rev. B {\bf 19}, 3593 (1979)

\bibitem{Grupp1997}
D. E. Grupp \etal, Science {\bf 276}, 392 (1997)

\bibitem{Fleury1968}
P. A. Fleury \etal, Phys. Rev. Lett. {\bf 21}, 16 (1968)

\bibitem{Sawaguchi1963}
E. Sawaguchi \etal, J. Phys. Soc. Jpn. {\bf 18}, 459 (1963)

\bibitem{Hoppler2008}
J. Hoppler \etal, Phys. Rev. B {\bf 78}, 134111 (2008)

\bibitem{Kalisky2013}
B. Kalisky \etal, Nature Materials {\bf 12}, 1091 (2013)

\bibitem{Honig2013}
M. Honig \etal, Nature Materials {\bf 12}, 1112 (2013)

\bibitem{Chang1972}
T.S. Chang \etal, J. Appl. Phys. {\bf 43}, 3591 (1972)

\bibitem{Muller1970}
K. A. M\"uller \etal, Solid State Commun. {\bf 8}, 549 (1970)

\bibitem{vanMechelen2010thesis}
D. van Mechelen,  PhD thesis, University of Geneva, (2010)

\bibitem{Chang1970}
T.S. Chang \etal, Appl. Phys. Lett. {\bf 17}, 254 (1970)

\bibitem{Wang2014}
Z. Wang \etal, PNAS {\bf 111}, 3933 (2014)

\bibitem{Annadi2013}
A. Annadi \etal, Nat. Commun. {\bf 4}, 1838 (2013)

\bibitem{Bottin2005}
F. Bottin \etal, Surf. Sci. {\bf 574}, 65 (2005)

\bibitem{Eglitis2008}
R. I. Eglitis \etal, Phys. Rev. B {\bf 77}, 195408 (2008)

\bibitem{Enterkin2010}
J.A. Enterkin \etal, Nature Mat. {\bf 9}, 245 (2010)

\bibitem{Chrosch1998}
J. Chrosch \etal, J. Phys.: Condens. Matter {\bf 10}, 2817 (1998)

\bibitem{Bernhard2004}
C. Bernhard \etal, Thin Solid Films {\bf 455}, 143 (2004)

\bibitem{Marsik2016}
P. Marsik \etal, Appl. Phys. Lett. {\bf 108}, 052901 (2016)

\bibitem{Aspnes1980}
D. E. Aspnes, J. Opt. Soc. Am. {\bf 70}, 1275 (1980)

\bibitem{Kozuka2007}
Y. Kozuka \etal, Phys. Rev. B {\bf 76}, 085129 (2007)

\bibitem{Hlinka2006}
J. Hlinka \etal, Phase Transit. {\bf 79}, 41 (2006)

\bibitem{Trautmann2004}
T. Trautmann \etal, J. Phys.: Condens. Matter {\bf 16}, 5955 (2004)

\bibitem{Petzelt2001}
J. Petzelt \etal, Phys. Rev. B {\bf 64}, 184111 (2001)

\bibitem{Rossle2013prl}
M. R\"ossle \etal, Phys. Rev. Lett {\bf 110}, 136805 (2013)

\bibitem{Yamanaka2000}
A. Yamanaka \etal, Europhys. Lett. {\bf 50}, 688 (2000)

\bibitem{Humlicek2000}
J. Humli\v{c}ek \etal, Phys. Rev. B {\bf 61}, 14554 (2000)

\bibitem{Fedorov1998}
I. Fedorov \etal, Ferroelectrics 208–209 {\bf 1}, 413 (1998)

\bibitem{Rossle2013thesis}
M. R\"ossle, PhD thesis, University of Fribourg, 2013

\bibitem{Rossle2013prb}
M. R\"ossle \etal, Phys. Rev. B {\bf 88}, 104110 (2013)

\bibitem{Guennuo2010}
M. Guennuo \etal, Phys. Rev. B {\bf 81}, 054115 (2010)

\bibitem{Berreman1963}
D.W. Berreman, Phys. Rev. {\bf 130}, 2193 (1963)

\bibitem{Dubroka2010}
A. Dubroka \etal, Phys. Rev. Lett. {\bf 104}, 156807 (2010)

\bibitem{Yazdi-Rizi2016}
M.Yazdi-Rizi \etal, Europhys. Lett. {\bf 113}, 47005 (2016)



\end{thebibliography}
\end{document}